\documentclass[prd]{revtex4}                                                   
\usepackage[pdftex]{graphicx}
\usepackage{wrapfig, epsfig}

\newcommand{\keyword}[1]{\textsf{\slshape #1}}

\begin{document}
\title{Dynamics of Dark Matter in Baryon-Radiation Plasma:\\
Perspectives using Meschersky equation}
\author{Himanshu kumar}
\email{hman_19@hotmail. com}
\author{Sharf Alam}
\affiliation{Jamia Millia Islamia,  New Delhi,  India}

\begin{abstract}
With an aim to argue for the truly collisionless nature of cold
dark matter between epochs of equality and recombination, we assume
a model, wherein strongly coupled baryon-radiation plasma ejects out
of small regions of concentrated cold dark matter without losing its
equilibrium.  We use the Meschersky equation to describe the dynamics of
cold dark matter in the presence of varying mass of strongly coupled
baryon-radiation plasma. 
Based on this model, we discuss the growth of perturbations in cold dark
matter both in the Jeans theory and in the expanding universe using
Newton's theory.  We see the effect of the perturbations in the cold
dark matter potential on the cosmic microwave background anisotropy that
originated at redshifts between equality and recombination i.e. 
$1100 < z < z_{eq}$.  Also we obtain an expression for the Sachs-Wolfe effect,
i.e. the CMB temperature anisotropy at decoupling in terms of the perturbations in
cold dark matter potential.  We obtain  similar solutions both in the static
and in the expanding universe, for epochs of recombination.  From this, we
infer about the time scale when the dark energy starts to dominate. 
\end{abstract}
\maketitle

\keyword{Meschersky equation, Cold Dark
matter, Collisionless, Baryon-Radiation plasma, Cosmic Microwave
background, Sachs-Wolfe effect}

\section{Introduction}
\label{sec:1 Intro}
Flat cosmological models with a mixture of ordinary baryonic matter, cold
dark matter, and cosmological constant(or quintessence) and a nearly
scale-invariant, adiabatic spectrum of density fluctuations are consistent
with standard inflationary cosmology.  They provide an excellent fit to
current observations on large scales($>>1$ Mpc).  Currently, the constitution
of the universe is 4\% baryons, 23\% dark matter and 73\% dark energy
\cite{Reid2010,Percival2010,Huff,Tinker2012,Larson2011,Dunkey2011}. 

In the standard hot Big Bang model, the universe
is initially hot and the energy density is dominated by radiation.  The
transition to matter domination occurs at $z\approx 10^{4}$.  In the
epochs after equality and before recombination, the universe remains
hot enough.  Thus the gas is ionized, and the electron-photon scattering
effectively couples the matter and radiation \cite{Peebles1970}.  At
$z\approx 1200$, the temperature drops below $\approx 3300$ K.  The
protons and electrons now recombine to form neutral hydrogen
and neutral helium.  This event is usually known as recombination
\cite{Peebles1968,Zeldovich1969,Seager2000}.  The
photons then decouple and travel freely.  These photons which
keep on travelling till present times are observed as the cosmic
microwave background(CMB).  The cold dark matter theory including
cosmic inflation is the basis of standard modern cosmology.  This
is favoured by the CMB data and the large scale structure data
\cite{PJE1982,Pagels1982}.  The CDM model is based on the assumption
that the mass of the universe now is dominated by dark matter,  which
is non-baryonic \cite{Tegmark2004,Spergel2007}.  Also it acts
like a gas of massive, weakly interacting(collisionless)particles
\cite{BondSzalay1983}.  They have negligibly small primeval
velocty dispersion.  Also they are electromagnetically neutral
\cite{Sigurdson2004,Mcdermott2011}.

The word Cold here means that
the ratio $\frac{T}{M}<\phi$, the gravitation potential, where T and M
represent the temperature and mass of the dark matter particle.  There is
remarkably good agreement between standard CDM models and the observed
power spectrum of Lyman $\alpha$ observers \cite{Croft1999}.  This rules
out the warm dark matter candidates.  The CDM model predicts the power
spectrum of the angular distribution of the temperature of the 3K
cosmic microwave background radiation and the flat Friedmann model.  A
low density CDM model with a density parameter of around 0. 3 to 0. 4
with cosmological constant actually matches all available data fairly
well \cite{Efstathiou1990,Ostriker2002}.  The stable CDM paradigms
predict all structure formation \cite{Peebles1980,White2010}. 

The
existence of clusters($\leq 50$ mpc) and groups of galaxies suggests
that the galaxy formation is due to the gravitational instability
of a spatially homogenous and isotropic expanding universe.  The
Perturbations of such a model have been first investigated by
\cite{Jeans1902,cambridgeunivpress1928}.  Then within the framework
of general relativity by \cite{Lifshitz1946,Khalatnikov1963}.  Also
it was studied in a newtonian model\cite{Bonnor1957}.  The results
obtained from relativistic theory were similar to the standard
Jeans theory.  In the early universe, radiation fixes the expansion
rate.  This is due to low density and self-gravity of dark matter
\cite{Guyot1970,Meszaros1974}.  After equality, the main contribution
to the gravitational potential is due to the cold dark matter.  

In the metric of the perturbed FRW universe, the main contribution to
the gravitational potential comes from an imperfect fluid, i. e.  the dark
matter.  We still consider that the difference $\psi - \phi $ is suppressed
compared to $ \psi $, atleast by the ratio of the photon mean free path to
the perturbation scale.  Here $\psi$ and $\phi$ are the newtonian potential
and the perturbations to the spatial curvature in a conformal Newtonian
gauge respectively(see eq. 48 sec~\ref{sec:2}).  We neglect the contribution
of the baryon density in realistic models,  where baryon contributes only
a small fraction of the total matter density.  At $ \eta > \eta_{eq}$, the
gravitational potential is mainly due to the perturbations in the cold
dark matter.  The potential of cold dark matter is time-independent both
for long wavelength and short wavelength perturbations.  This is because
it is highly non-relativistic.  There is a strong coupling between the
baryon and the radiation in the plasma.  So, we treat it as a single perfect
fluid for low baryon densities.  We neglect the non-diagonal components
in the energy- momentum tensor of dark matter.  This is because for
very small photon mean free paths  $\psi$ is same as $\phi$.  Therefore, 
we treat it as a single perfect fluid for many epochs after equality
upto the recombination.  It is only after recombination,  that the baryon
density starts increasing.  This is after the primordial nucleosynthesis
of hydrogen and helium is complete.  The radiation after decoupling from
matter at very late recombination epochs evolves separately.  The CMB
anisotropy of epochs between equality and recombination gives important
information about the perurbations in the cold dark matter i. e.   the
modes that enter the horizon before recombination. 

In our model, the strongly coupled baryon-radiation plasma ejects out
of a concentrated region of cold dark matter.  Based on this model, we
study the dynamics of the cold dark matter after equality and upto many
epochs near recombination.  Note that this sort of flow assumes that the
baryon -radiation plasma ejects out without losing its equilibrium in
the presence of cold dark matter.  This is possible only if we assume
that the dark matter is truly collisionless.  This assumption can also
accomodate a self-interacting dark matter at very small scales, which can
transfer energy and momentum to the outer core \cite{Spergel2000}.  The
strongly-coupled baryon -radiation plasma is only under the influence
of the potential of cold dark matter at these epochs.  We assume that the
number of photons per Cold dark matter is initially spatially uniform
on supercurvature scales(wavelengths greater than $ H^{-1}$).  During
this time, the matter and radiation densities vary in space.  In other
words we consider adiabatic perturbations.  As the universe expands, the
inhomogeneity scale becomes smaller than the curvature scale.  Thus the
components move with respect to one another and the entropy of photon
per cold dark matter particle varies spatially.  In contrast the entropy
per baryon remains spatially uniform on all scales.  This is until the
baryons decouple from radiation. 

We use Meshchersky equation to study the dyanmics of the strongly
coupled baryon- radiation plasma after the cold dark matter starts
to dominate.  This occurs after epochs of equality.  In this model, we
imagine a flow of the baryon - radiation plasma across regions of highly
concentrated non -relativistic cold dark matter.  The baryon -radiation
plasma can be assumed to have an ejection velocity with respect to the
cold dark matter after equality.  We do not disturb the equilibrium of
the ejecting baryon-radiation plasma, even in the presence of cold dark
matter.  This we do to argue in favour of truly collisionless nature of
cold dark matter.  We ensure this by a specific assumption(see eq. 27
sec. ~\ref{sec:2}).  This is the time,  when the radiation densities
rapidly start to decrease.  This is due to its scaling proportional to
$a^{-4}$, which is $a^{-3}$ for the cold dark matter density.  On the basis
of this model,  we  study the adiabatic perturbations between epochs of
equality and recombination.  This we do, both in the Jean's theory and in
an expanding universe in Newtonian theory.  We then find out the effect
of perturbations in cold dark matter potential on the anisotropy in
temperature of radiation at these epochs.  This we do for modes greater
than the curvature scales.  These are the modes which enter horizon
before recombination. 

The paper proceeds as follows.  In section 2, we discuss the Dynamics
of Cold dark matter in Baryon-radiation plasma using the Meshchersky
equation in the Jeans theory.  Then we discuss the adiabatic perturbations
in the scope of this model.  We also deduce an equation which represents
how the anisotropy in the temperature of radiation is affected by the
perturbations in the cold dark matter in the epochs between equality
and recombination.  We derive an expression for the Sachs-Wolfe effect.  In
Section 3, in the scope of our model, we discuss the dynamics of the above
mentioned scenario in the expanding universe in the Newtonian theory.  We
show how the anisotropy in the temperature of radiation is affected
by the perturbations in Cold dark matter potential,  in an expanding
universe.  Also we evaluate an expression for the Sachs-Wolfe effect in
the expanding universe.  We then write the equation for the evolution of
cold dark matter perturbations with time. 

\section{Gravitational Instability: Jeans theory Representation}
\label{sec:2}
In Jeans theory we consider a static,  non-expanding universe.  Also we
assume a homogeneous, isotropic background with constant time -independent
matter density \cite{Jeans1902,cambridgeunivpress1928}.  This
assumption is in  obvious contradiction to the hydrodynamical equations.  In
fact, the energy density remains unchanged only if the matter is at
rest and the gravitational force, $F \propto \bigtriangledown\phi $
vanishes.  This inconsistency can in principle be avoided if we consider
a static Einstein universe, where the gravitational force  of the matter
is compensated by the antigravitational force of an appropriately chosen
cosmological constant. 
We consider the fundamental equation of dynamics of a mass point with variable mass.  This is also referred to as the Meshchersky equation. 
\begin{equation}
 m\frac{\vec{dv}}{dt} = F + \frac{dm}{dt}\vec{u} 
\end{equation}
It should be pointed out that in an inertial frame,  $\vec{F}$ is
interpreted as the force of interaction of a given body with surrounding
bodies.  The last term          $\frac{dm}{dt}\vec{u}$ is referred to as
the reactive force.  This force appears as a  result of the action that
the added or separated mass exerts on a given body.  If mass is added
, then the $\frac{dm}{dt}>0$ coincides with the vector $\vec{u}$.  If mass
is separated, $\frac{dm}{dt}< 0$,  and the vector $\vec{R}$ is oppositely
directed to the vector $\vec{u}$. 

 We consider a fixed volume element $\Delta V$ in Euler (non-co-moving)
 co-ordinates $\vec{x}$.  After equality, when the cold dark matter starts
 to dominate in small regions of space, we write:
\begin{equation}
\Delta\, M_{d}\frac{\vec{v_{d}}}{dt}= -\Delta M_{d}\cdot \nabla \phi - \nabla p_{b\gamma}\cdot \Delta V
- \frac{dM_{b\gamma}}{dt}\vec{u}
\end{equation}
Here the first term on L. H. S represents the acceleration in the mass of
cold dark matter of mass $\Delta M_{d}$ and $\vec{v_{d}}$ is the velocity
of dark matter element.  The first term on the R. H. S.  represents the
gravitational force on the cold dark matter.  The second term on right
is for the force due to the pressure of the baryon -radiation plasma
\cite{Mukhanov2005}.  The last term on R. H. S.  is the reactive force
on the cold dark matter due to the ejection of baryon - radiation plasma
from regions dominated by cold dark matter. This last force arises only
due to the model that we assume here.  Here $p_{b\gamma}$ is the pressure
of the baryon-radiation plasma and $\vec{u}$ is the ejection velocity of
the Baryon - radiation plasma with respect to the concentrated region of
cold dark matter.  Note that in (eq. 2), we do not take the pressure of
cold dark matter into account.  This is because, the cold dark matter
is highly non - relativistic.  Therefore,  we neglect its pressure.
We assume that after equality, the strongly coupled Baryon - radiation
plasma starts to rapidly decouple from the matter.  The matter at these
epochs is predominantly Cold Dark matter.

After Equality, in regions dominated by highly non - relativistic Cold dark matter, we write the continuity equation for the ejection of baryon- radiation plasma complex :
\begin{equation}\
\frac{dM_{b\gamma}}{dt}=\int _{\Delta V} \frac{\partial \varepsilon_{b\gamma}}{dt}dV
\end{equation}
where $M_{b\gamma}$ is the ejecting mass of baryon-radiation plasma and
$ \varepsilon_{b\gamma}$
is the energy density of baryon radiation plasma in the concentrated
region of cold dark matter, from where it is ejecting out.  In this model, 
we assume that     the ejection of baryon -radiation plasma out of
a concentrated region of heavy Cold dark matter does not disturb the
equilibirium of the dark matter.  This  we can assume  only because of the
truly collisionless nature of Cold dark matter.  This is the fundamental
assumtion of this model.  We neglect the flux of the Cold dark matter out
of a region of volume $\Delta V$ in the time that the baryon -radiation
plasma flows out of this region.  The rate of flow is entirely determined
by the flux of the baryon - radiation plasma and   we write :
\begin{equation}
\frac{dM_{b\gamma}}{dt}= \int _{\Delta V} \frac{\partial \varepsilon_{b\gamma}}{dt}dV = - \int _{\Delta V}\vec{\nabla}\cdot \varepsilon _{b\gamma}\vec{u}dV
\end{equation}
 We  write (Eq. 2) as:
\begin{equation}
\varepsilon_{d}\Delta V\frac{d\vec{v_{d}}}{dt}= -\varepsilon_{d}\Delta V\nabla\phi - \nabla p_{b\gamma}\Delta V - \frac{\partial\varepsilon_{b\gamma}}{\partial t}\Delta V\vec{u}
\end{equation}
where $\varepsilon_{d}$ is the energy density of cold dark matter, and $
\phi$ is the gravitational potential of cold dark matter. 
\begin{equation}
\varepsilon_{d}[\frac{\partial\vec{v_{d}}}{\partial t} + \vec{v_{d}}\cdot \nabla\vec{v_{d}}]= -\varepsilon_{d}\nabla\phi -\nabla p_{b\gamma} - \frac{\partial\varepsilon_{b\gamma}}{\partial t}\vec{u}
\end{equation}
Now using the Jeans theory, we introduce small perturbations about the
equilibrium values of variables, W. Bonnor\cite{Bonnor1957}:
\begin{equation}
\varepsilon_{d}(\vec{x}, t)=\varepsilon_{do} + \delta\varepsilon_{d}(\vec{x}, t)
\end{equation}
\begin{equation}
\vec{v_{d}}(\vec{x}, t)= \vec{v_{do}} + \delta\vec{v_{d}}(\vec{x}, t)= \delta\vec{v_{d}}(\vec{x}, t)
\end{equation}
\begin{equation}
\phi(\vec{x}, t)= \phi_{o}+\delta\phi (\vec{x}, t)
\end{equation}
\begin{equation}
\vec{u}(\vec{x}, t)= \vec{u_{o}} + \delta\vec{u}(\vec{x}, t)
\end{equation}
where $\vec{v_{do}}<<c$, the speed of light, $\vec{u_{o}}\neq 0$ and $\delta\varepsilon_{d}<< \varepsilon_{do}$ 
\begin{equation}
p_{b\gamma}(\vec{x}, t)= p_{b\gamma}(\varepsilon_{b\gamma o} + \delta\varepsilon_{b\gamma}, S_{o} + \delta S) = p_{b\gamma o} + \delta p_{b\gamma}(\vec{x}, t)
\end{equation}
\begin{equation}
S(\vec{x}, t)= S_{o} + \delta S(\vec{x}, t)
\end{equation}
where S is the entropy of cold dark matter element.  Also we write:
\begin{equation}
\delta p_{b\gamma} = c_{s}^{2}\delta\varepsilon_{b\gamma} + \sigma\delta S
\end{equation}
where $c^{2}_{s}$ is the speed of sound.  Neglecting Dissipation, we write:
\begin{equation}
\frac{dS}{dt} = \frac{\partial S}{\partial t} +  \vec{v_{d}}\cdot \vec{\nabla}S
\end{equation}
\begin{equation}
\nabla^{2}\phi = 4\pi G\varepsilon_{d}
\end{equation}
From (eq. 4) we write:
\begin{equation}
\frac{\partial\varepsilon_{b\gamma}}{\partial t} + \vec{\nabla}\cdot \varepsilon_{b\gamma}\vec{u} = 0
\end{equation}
We substitute the values from (eq. 7)to(eq. 15)in(eq. 16)and(eq. 6) to get:
\begin{equation}
\frac{\partial\delta\varepsilon_{b\gamma}}{\partial t} + \varepsilon_{b\gamma o}\vec{\nabla}\cdot \delta\vec{u} + \vec{\nabla}\cdot (\delta\varepsilon_{b\gamma}\vec{u_{o}}) = 0
\end{equation}
\begin{equation}
\varepsilon_{do}\frac{\partial\delta\vec{v_{d}}}{\partial t} = - \delta\varepsilon_{d}\nabla\phi_{o} - \nabla(c_{s}^{2}\delta\varepsilon_{b\gamma} + \sigma\delta S_{b\gamma}) - \frac{\partial\varepsilon_{b\gamma 0}}{\partial t}\delta\vec{u}  - \frac{\partial\delta\varepsilon_{b\gamma}}{\partial t}\vec{u_{o}}
\end{equation}
\begin{equation}
\nabla^{2}\delta\phi = 4\pi G\delta\varepsilon_{d}
\end{equation}
\begin{equation}
\frac{d\delta S}{dt} = \frac{\partial\delta S}{dt} + (\delta\vec{v_{d}}\cdot \nabla)S_{o} = 0 
\end{equation}
We now take the divergence of (eq. 18)to get:
\begin{eqnarray}
\varepsilon_{do}\frac{\partial \vec{\nabla}\cdot \delta\vec{v_{d}}}{\partial t} = - (\vec{\nabla}\cdot \delta\varepsilon_{d})\nabla\phi_{o} - \delta\varepsilon_{d}(4\pi G\varepsilon_{do}) 
-c_{s}^{2}\nabla^{2}\delta\varepsilon_{b\gamma}\nonumber \\
 - \sigma\nabla^{2}\delta S  - \vec{\nabla}\cdot \lbrace\frac{\partial(\varepsilon_{b\gamma o}\delta\vec{u})}{\partial t} - \varepsilon_{b\gamma o}\frac{\partial\delta\vec{u}}{\partial t}\rbrace - \vec{\nabla}\cdot \lbrace \frac{\partial (\delta\varepsilon_{b\gamma }\vec{u_{o}})}{\partial t} - \delta\varepsilon_{b\gamma}\frac{\partial \vec{u_{o}}}{\partial t}\rbrace  
\end{eqnarray}
Now using (eq. 17) and (eq. 19) we get :
\begin{equation}\frac{\partial^{2}\delta\varepsilon_{b\gamma}}{\partial t^{2}} - c_{s}^{2}\nabla^{2}\delta\varepsilon_{b\gamma} - 4\pi G \varepsilon_{do}\delta\varepsilon_{d} - (\vec{\nabla}\cdot \delta\varepsilon_{d})\nabla\phi_{o}
 -\varepsilon_{do}\frac{\partial(\vec{\nabla}\cdot \delta\vec{v_{d}})}{\partial t}  + \vec{\nabla}\cdot \lbrace\varepsilon_{b\gamma o}\frac{\partial\delta\vec{u} }{\partial t} + \delta\varepsilon_{b\gamma}\frac{\partial\vec{u_{o}}}{\partial t}\rbrace = \sigma\nabla^{2}\delta S
\end{equation}
In a static universe the total energy remains constant.  So for small inhomogeneities we write:
\begin{equation} \delta\varepsilon_{b\gamma} = -\delta\varepsilon_{d}\end{equation}
Thus (eq. 22) becomes:
\begin{equation}\frac{\partial^{2}\delta\varepsilon_{b\gamma}}{\partial t^{2}} - c_{s}^{2}\nabla^{2}\delta\varepsilon_{b\gamma} + 4\pi G\varepsilon_{do}\delta\varepsilon_{b\gamma} +(\vec{\nabla}\cdot \delta\varepsilon_{b\gamma})\nabla\phi_{o} 
-\varepsilon_{do}\frac{\partial\vec{\nabla}\cdot \vec{v_{d}}}{\partial t} + \vec{\nabla}\cdot \lbrace\varepsilon_{b\gamma o}\frac{\partial\delta\vec{u}}{\partial t} + \delta\varepsilon_{b\gamma}\frac{\partial\vec{u}}{\partial t}\rbrace
= \sigma\nabla^{2}S(x)
\end{equation}
For strongly -coupled baryon - radiation plasma before recombination, with low baryon densities we  write :
\begin{equation}\varepsilon_{b\gamma}+ p_{b\gamma}=\varepsilon_{b}
+ \frac{4}{3}\varepsilon_{\gamma} = \frac{4}{9c_{s}^{2}}\varepsilon_{\gamma}\end{equation}
Thus we get:
\begin{equation}
\delta\varepsilon_{b\gamma} + \delta p_{b\gamma} = \frac{4}{9c_{s}^{2}}\delta\varepsilon_{\gamma}
\end{equation}
The baryon - radiation plasma is only affected by the gravitational
potential of the cold dark matter in these epochs.  We argue that the
baryon-radiation plasma ejecting out of concentrated regions of cold
dark matter is always in equilibrium, even in the presence of cold
dark matter.  This is possible only if the cold dark matter is truly
collisionless.  So for epochs between equality and recombination, we write:
\begin{equation}
 \frac{\partial p_{b\gamma}}{\partial T} = \frac{\varepsilon_{b\gamma} + p_{b\gamma}}{T}\end{equation}
The above equation is of profound importance in this model.  This
is because,  it ensures that the equilibrium of the ejecting baryon
-radiation plasma is not disturbed, even when it flows out of concentrated
regions of cold dark matter.  This we can write, because the distribution
function of pressure of the baryon-radiation plasma depends only on
$\frac{E}{T}$.  We assume here that the chemical potential is much smaller
than the temperature.  So from small perturbations about the equilibrium
values of variables $p_{b\gamma}$ and$\varepsilon_{b\gamma}$ in the
(eq. 27), about a fixed temperature T,  we get, 
   \begin{equation} \frac{\delta p_{b\gamma}}{\delta T} = \frac{4\delta\varepsilon_{\gamma}}{9c_{s}^{2}T} 
   \end{equation}
   Thus we write:
   \begin{equation}
   \delta\varepsilon_{b\gamma} = \frac{4\delta\varepsilon_{\gamma}}{9c_{s}^{2}}\lbrace1 - \frac{\delta T}{T}\rbrace \end{equation}
   Here $\frac{\delta T}{T}\equiv\Theta$, which represents the
   anisotropy in temperature of radiation for small baryon densities
   after equality.  For small perturbations in cold dark matter, using
   (eq. 29) in (eq. 24), we write:
  
 \begin{eqnarray}
 \frac{4}{9c_{s}^{2}}\frac{\partial^{2}\delta\varepsilon_{\gamma}\lbrace 1 - \Theta\rbrace}{\partial t^{2}} - \frac{4}{9}\nabla^{2}\lbrace\delta\varepsilon_{\gamma}\lbrace
1 -\Theta\rbrace\rbrace +\nabla^{2}\phi_{o}\frac{4}
{9c_{s}^{2}}\delta\varepsilon_{\gamma}\lbrace 1 -\Theta\rbrace +\frac{4}{9c_{s}^{2}}\vec{\nabla}\cdot \lbrace\delta\varepsilon_{\gamma}(1 - \Theta)\rbrace\nabla\phi_{o} \nonumber\\
-\varepsilon_{do}\frac{\partial(\vec{\nabla}\cdot \delta\vec{v_{d}})}{\partial t}+\vec{\nabla}\cdot \lbrace\varepsilon_{b\gamma o}\frac{\partial\delta\vec{u} }{\partial t}\rbrace +\frac{4}{9c_{s}^{2}}\vec{\nabla}\cdot \lbrace\delta\varepsilon_{\gamma}(1 -\Theta)\frac{\partial\vec{u_{o}}}{\partial t}\rbrace =\sigma\nabla^{2}\delta{S(x)}
\end{eqnarray} 

For small perturbations in Cold dark matter,  we neglect the term $\vec{\nabla}\cdot \delta\vec{v_{d}} $.  This is due to cold dark matter being highly non -relativistic.  Also we neglect $ \vec{\nabla}\cdot \frac{\partial\delta\vec{u}}{\partial t} $.  This is due to the fact, that in this model,  we assume that the ejecting velocity of baryon - radiation plasma does not suffer interactions and collisions.  This is because of dark matter being truly collisionless.  So, we further write: \begin{eqnarray}
\frac{4}{9c_{s}^{2}}\frac{\partial^{2}\delta\varepsilon_{\gamma}\lbrace 1 - \Theta\rbrace}{\partial t^{2}} - \frac{4}{9}\nabla^{2}\lbrace\delta\varepsilon_{\gamma}\lbrace
1 - \Theta\rbrace\rbrace +
 4\pi G\varepsilon_{do}\frac{4}{9c_{s}^{2}}\delta\varepsilon_{\gamma}\lbrace 1 - \Theta\rbrace \nonumber \\ +\frac{4}{9c_{s}^{2}}\vec{\nabla}\cdot \lbrace\delta\varepsilon_{\gamma}(1 -\Theta)\rbrace\nabla\phi_{o} +\frac{4}{9c_{s}^{2}}\vec{\nabla}\cdot \lbrace\delta\varepsilon_{\gamma}(1-\Theta)\frac{\partial\vec{u_{o}}}{\partial t}\rbrace
 =\sigma\nabla^{2}\delta{S(x)}
 \end{eqnarray}
 In the above equation,  we  see the contribution of divergence in the
 anisotropy of radiation at epochs between equality and recombination.  At
 epochs near recombination, the strongly coupled baryon-radiation plasma
 starts rapidly to decouple from the cold dark matter.  During these
 epochs, therefore the actual velocity of the baryon -radiation plasma
 does not change appreciably due to the gravitational force of cold dark
 matter.  During these epochs, the rate of increase of ejection velocity
 is only due to gravitation potential of cold dark matter.  There is no
 force due to the pressure of the baryon -radiation plasma.  Therefore
 we write:\begin{equation}\frac{\partial\vec{u_{o}}}{\partial t} =
 -\nabla\phi_{o}\cdot \end{equation}
 We can argue that due to high baryon densities, the pressure of
 the baryon-radiation plasma vanishes at these epochs.  But the baryon
 densities can never be very high.  This is because it will hinder hydrogen
 nucleosynthesis after recombination.  Thus it has to be assumed  that
 the pressure of the  baryon - radiation plasma   vanishes not due to
 very high baryon densities, but due to a strange form of energy(dark
 energy)with negative pressure, which has started to dominate near
 recombination.  Therefore, the acceleration in the ejection velocity will
 be primarily due to the gravitational forces So, we write:
 \begin{equation}\vec{u_{o}}= \vec{v_{b\gamma o}} -\vec{v_{do}}\end{equation}
  \begin{equation} \dot{\vec{u_{o}}} = \dot{\vec{v_{b\gamma o}}} - \dot{\vec{v_{do}}}\end{equation}
 \begin{equation} \dot{\vec{v_{b\gamma o}}}= 0
 \end{equation}
 The above (eq. 35) shows that the actual velocity of the baryon-radiation
 plasma stops to increase at epochs, where it is almost about to decouple
 from cold dark matter.  Therefore, at these epochs the accleration of cold
 dark matter reverses its sign and we write:
 \begin{equation}
 \dot{\vec{v_{do}}}=\nabla\phi_{o}
 \end{equation}
 With these Assumptions  the (eq. 31) reduces to:
 \begin{equation}\frac{\partial^{2}}{\partial t^{2}}\lbrace\delta\varepsilon_{\gamma}
 1 - \Theta\rbrace{\partial t^{2}} -c_{s}^{2}\nabla^{2}\lbrace\delta\varepsilon_{\gamma}(1 - \Theta)\rbrace +4\pi G\varepsilon_{do}\lbrace\delta\varepsilon_{\gamma}(1 - \Theta)\rbrace
 =\frac{9c_{s}^{2}}{4}\sigma\nabla^{2}\delta{S(x)}
\end{equation}
Considering adiabatic perturbations:
\begin{equation}\delta S = 0\end{equation}
We use:
\begin{equation}\delta\varepsilon_{\gamma}(1 - \Theta)(\vec{x}, t) = \int\delta\varepsilon_{\gamma k}(t)(1 - \Theta)_{k}(t)e^{i\vec{k}\cdot \vec{x}} \frac{d^{3}k}{(2\pi)^{3/2}}\end{equation} 
Then  we write:
 \begin{equation}
 \delta\varepsilon_{\gamma k}(t)(1 - \theta)_{k}(t) = y_{k}(t)
 \end{equation}
The (eq. 37) now reduces to:
\begin{equation}
\ddot{y_{k}}(t) + k^{2}c_{s}^{2}y_{k}(t) + 4\pi G\varepsilon_{do}y_{k}(t)=0\end{equation}
\begin{center}or\end{center}
\begin{equation}
y^{. . }_{k}(t) +(k^{2}c_{s}^{2} + 4\pi G\varepsilon_{do})y_{k}(t)=0
\end{equation}
The above equation has two solutions:$y_{k}\propto e^{\pm i\omega(t)}$, where
\begin{equation}
\omega(k)=\sqrt{k^{2}c_{s}^{2} + 4\pi G\varepsilon_{do}}=\sqrt{k^{2}c_{s}^{2} - 4\pi G\varepsilon_{b\gamma o}(1 - \frac{\varepsilon_{o}}{\varepsilon_{b\gamma o}})}
\end{equation}
Here the Jeans length is :
\begin{equation}
\lambda_{J} =\frac{2\pi}{k_{J}} = c_{s}\sqrt{\frac{\pi}
{G\varepsilon_{b\gamma o}}\lbrace 1 -
\frac{\varepsilon_{o}}{\varepsilon_{b\gamma o}}\rbrace}
\end{equation}
Thus $\omega(k)$ is real for $\lambda < \lambda_{J}$ or $k>k_{J}$, where \begin{equation}
k_{J} = \frac{2\pi}{c_{s}}\sqrt{\frac{G\varepsilon_{b\gamma o}\lbrace 1 - \frac{\varepsilon_{o}}{\varepsilon_{b\gamma o}}\rbrace}{\pi}}
\end{equation}
when 
\begin{equation} k^{2}c_{s}^{2} > 4\pi G\varepsilon_{b\gamma o}\lbrace 1 - \frac{\varepsilon_{o}}{\varepsilon_{b\gamma o}}\rbrace
\end{equation}
Where $\varepsilon_{o}$ is the total equilibrium value of energy density and $\varepsilon_{b\gamma o}$ is the equilibrium energy density of baryon-radiation plasma, at these epochs.  We interpret from above,  that all the modes near recombination are real.  This is because the value of $\frac{\varepsilon_{o}}{\varepsilon_{b\gamma o}}$ starts increasing rapidly  near recombination.  This is due to the decoupling of baryon - radiation plasma from cold dark matter.  This explains the maximum frequency of fluctuations in the CMB temperature anisotropy at red -shifts of recombination.  For $ \lambda < \lambda_{J}$ the solutions are:
\begin{equation}
y_{k}(\vec{x}, t)\propto \sin(\omega t + kx +\alpha)
\end{equation}
For $\lambda < \tau_{\gamma} < \lambda_{J}$, where $\tau_{\gamma}$
stands for the mean free path of free- streaming photons after
equality, free-streaming becomes important.  Free-streaming refers
to the propagation of photons without scattering.  We write the
equation for free-streaming photons in the conformal Newtonian gauge
\cite{Scott2003}.  The gravitational potential is primarily due to the
cold dark matter.  We argue that the photons  can still be described by
the equilibrium distribution functions.  This is because the cold Dark
matter is collisionless.  Therefore,  their mutual interactions will not
disturb the equilibrium of the ejecting baryon-radiation plasma.  Also, we
treat the baryon-radiation plasma as free-streaming photons for low
baryon-densities.  We use the metric below:
\begin{equation}
 ds^{2}= -1 -2\psi(\vec{x}, t)dt^{2} + a^{2}\delta_{ij}(1 + 2\phi(\vec{x}, t))dx^{i}dx^{j}
 \end{equation}
where $\psi$ corresponds to the Newtonian potential and $\phi$ is the
perturbation to the spatial curvature. 
The equation for free- streaming photons is:
\begin{equation}
\frac{1}{p}\frac{dp}{dt}= -H -\frac{\partial\phi}{\partial t} - \frac{\hat{p}^{i}}{a}\frac{\partial \psi}{\partial x^{i}}
\end{equation}
\begin{equation}
\psi =\phi
\end{equation}
We can make the above assumption, for epochs between equality and
recombination.  This is because after Equality, and before recombination,  the
dominant contribution to the potential is due to cold dark matter.  So, we
neglect the non-diagonal components of the energy-momentum tensor of
the baryon-radiation plasma.  Recall that the difference $\psi - \phi$ is
suppressed compared to $\psi$, atleast by the ratio of the photon mean
free path to the perturbation scale.  Thus we assume it to be a single
perfect fluid.  In a static Einstein universe, we neglect the terms due
to the Hubble parameter $H$.  At scales of the order of mean-free path
of the free-streaming phtons in the baryon-radiation plasma, i. e.  of
the order of $\tau_{\gamma}$, we neglect the spatial inhomogeneities
$\frac{\partial\phi}{\partial\vec{x}}$.  We therefore write:
\begin{equation}
\frac{1}{p}\frac{dp}{dt} = -\frac{\partial\phi}{\partial t}
\end{equation} 
Using the assumption of (eq. 50), we write:
\begin{equation}
\frac{1}{p}\frac{dp}{dt} =-\frac{\partial\psi}{\partial t}
\end{equation}
Recall that the Newtonian potential $\psi$  corresponds to the dark
matter potential $\phi_{do}$.  This $\phi_{do}$ remains constant between
epochs of equality and recombination.  Thus we write:
\begin{equation}
\log p = -\psi \end{equation}
 \begin{equation}\log p = -\phi_{do}\Rightarrow p=e^{-\phi_{do}}\Rightarrow\delta p=-e^{-\phi_{do}}\delta\phi_{d}\end{equation} 
The momentum per unit volume of radiation at epochs of decoupling is
equivalent to radiation pressure.  Also, if the photons,  just at the epochs
of decoupling are in thermal equilibrium, and that it has same average
energy associated with each independent degree of freedom, we write:
\begin{equation}
\delta p_{\gamma}=\frac{1}{3}\delta p
\end{equation}
where $\delta p_{\gamma}$ is the increment in radiation pressure
associated with each independent degree of freedom at decoupling.  Since
for a unit volume decoupling photons' energy density is same as radiation
pressure per degree of freedom, we write:
\begin{equation}
\delta\varepsilon_{\gamma}= -\frac{1}{3}e^{-\phi_{do}}\delta\phi_{d}
\end{equation}
We use (eq. 40 and (eq. 47) to write:
\begin{equation}\delta\varepsilon_{\gamma k}(1 - \Theta)(\vec{x}, t) = A\sin(\omega t + kx + \alpha)\end{equation} 
where \begin{equation}\omega = \sqrt{k^{2}c_{s}^{2} - 4\pi G\varepsilon_{b\gamma o}(1 - \frac{\varepsilon_{o}}{\varepsilon_{b\gamma o}})}\end{equation}
Therefore,  for epochs of decoupling, we write: \begin{equation}
\frac{-e^{-\phi_{do}}\delta\phi_{d}}{3}\lbrace 1 - \Theta\rbrace  = A\sin(\omega t + kx + \alpha) \end{equation} \begin{center}or\end{center}
\begin{equation}
\lbrace\Theta -1\rbrace = \frac{3e^{\phi_{do}}}{\delta\phi_{d}}A\sin(\omega t + kx + \alpha) \end{equation}
 From the above equation we can write:
 \begin{equation}
 \frac{\delta\phi_{d}}{3}=\lbrace\theta
+\phi_{d}(1+\theta)^{2}\rbrace
\end{equation}  
  Therefore, we write:
  \begin{equation}
  \theta \approx\frac{\delta\phi_{d}}{3}
  \end{equation}
 The above result is the same as that predicted by Sachs-Wolfe.  Three
 types of effects(due to fluctuations in density, velocities
 and potential)simultaneously contribute to the CMB temperature
 anisotropy.  The fluctuations that matter at scales beyond $1^{\circ}$
 are those in the gravitational potential $\delta\phi_{d}$(Sachs-Wolfe
 effect)\cite{Sach1967}. 

The above equation represents the anisotropy in the CMB temperature, for
epochs near recombination, for regions where sufficient primordial helium
synthesis takes place, even before recombination, or the dark energy has
started to dominate.  Recall that in assumption of(eq. 32), we argue that
the force due to pressure of baryon-radiation plasma vanishes near
recombination epochs, and that it can be only due to the fact that the
dark energy with negative pressure has started to dominate.  The value of
$\phi_{do}$ is constant.  This is because the potential of cold dark matter
remains constant for many epochs between equality and recombination.  Thus
in a static universe, with this model,  we see that the dominant component
in CMB temperature anisotropy fluctuations is near recombination.  This
is because,  at epochs near recombination, the perturbations in the cold
dark matter potential are very small.  This is because in this model, the
decoupling of the baryon - radiation plasma from concentrated regions
of cold dark matter, near epochs of recombination is almost near
completion. 

We now write the CMB fluctuations for supercurvature modes i. e modes with
$\lambda >> \lambda_{J}$, and for regions where sufficient primordial
helium synthesis takes place, even before recombination, i. e.  the
baryon-densities are high.  We neglect the effect of gravity at these
epochs.  We can do so because at late recombination epochs, the pressure of
baryon-radiation plasma is low, and the negative pressure of a strange form
of energy(dark energy), which starts to dominate at these epochs,  cancels
the effect of forces due to low pressure baryon-radiation plasma and
that of gravity. we therefore write:\begin{equation}y_{k}\propto e^{\pm
ikc_{s}t}\end{equation}
Also for $\lambda >>\lambda_{J}$, free-streaming of photons is no longer
relevant.  Therefore the scattering of photons will dilute the anisotropy
to a large extent. 
For late      recombination epochs, when the radiation has almost decoupled
from matter,  we write:
  \begin{equation}c_{s}^{2} = \frac{1}{3}\Rightarrow y_{k}\propto e^{\pm \frac{ikt}{\sqrt{3}}}\end{equation}

We have neglected gravity here again, because at these epochs, the effect
of dark energy,  which had started to dominate from some earlier epochs is
to cancel the forces due to gravity and the pressure of baryon-radiation
plasma.  This is possible even in regions of lower baryon densities, which
has higher pressure than the regions of higher baryon densities.  This is
only because the dark energy with negative pressure had been dominating
from some earlier recombination epochs.  The above equation shows that
the frequency of fluctuations in CMB temperature anisotropy spectrum
of supercurvature modes(modes which enter the horizon very early
near recombination epochs),  at late recombination,  remain constant
till today.  This is valid both for regions of low or higher baryon
densities(where sufficient primordial helium nucleosynthesis takes place
before recombination).  This is because at very late recombination epochs
when the radiation has almost decoupled fully from matter, the speed of
sound approaches a constant value of $\frac{1}{3}$. 
The effect of scattering of photons will dilute this anisotropy in
supercurvature modes.  So, it is of not much cosmological significance. 

\section{Instability in Expanding Universe: Newtonian theory}
\label{sec:3}

Using our model (see Sec. ~\ref{sec:1 Intro}), we study the same
scenario, i. e.  of cold dark matter in the presence of strongly coupled
baryon -radiation plasma at epochs between equality and recombination.  Here
we use Newtonian theory of expanding universe \cite{Bonnor1957}.  We
treat the dynamics between cold dark matter and the baryon - radiation
plasma,  again in the framework of the Meshchersky equation.  We assume
that the strongly coupled baryon -radiation plasma is in the presence
of a gravitational potential.  This potential,  is only due to cold dark
matter, at epochs after equality and before recombination.  We neglect
the non-diagonal components of the energy-momentum tensor of the cold
dark matter.  This is because, the difference $\psi - \phi$ is suppressed
as compared to $\psi$,  atleast by the ratio of the mean free path to the
perturbation scale.  We consider dark matter as highly non -relativistic
fluid compared to the baryon - radiation plasma.  We consider epochs when
the Dark matter has already started clustering.  We neglect the rate of
flow of cold dark matter out of a given region of space, as compared to
the baryon-radiation plasma.  Therefore we write:
\begin{equation}
  \frac{\partial\varepsilon_{b\gamma}}{\partial t} + \vec{\nabla}\cdot (\varepsilon_{b\gamma}\vec{u}) = 0\end{equation}
where $\vec{u}$ is the relative velocity of the baryon - radiation
plasma with respect to the cold dark matter.  We write the Meshchersky
equation for cold dark matter with baryon radiation plasma,  ejecting
out of concentrated regions of space dominated  by cold dark matter:
  \begin{equation}
    \varepsilon_{do}\lbrace\frac{\partial\vec{v_{do}}}{\partial t}  + (\vec{v_{do}}\cdot \nabla)\vec{v_{do}}\rbrace = -\varepsilon_{do}\nabla\phi_{o} -\nabla p_{b\gamma} -\frac{\partial\varepsilon_{b\gamma o}}{\partial t}\vec{u_{o}} \end{equation} In an expanding flat, Isotropic and homogenous universe, we write:
 \begin{equation} \vec{v_{b\gamma}}=\vec{v_{b\gamma o}}(t); \varepsilon_{b\gamma}= \varepsilon_{b\gamma o}(t); 
 \vec{u_{o}}= H(t). \vec{x} ;  \varepsilon_{do} =\varepsilon_{do}(t);
\vec{v_{do}}=\vec{v_{do}(t)}\end{equation}
 Therefore we write:
 \begin{equation}
 \frac{\partial\varepsilon_{b\gamma o}(t)}{\partial t} + \varepsilon_{b\gamma o}(t)\vec{\nabla}\cdot \vec{u_{o}} =0 
 \end{equation}
We first discuss for scales $>>H^{-1}$ i. e the curvature scale.  In this
case,  we neglect the velocities of Cold dark matter particles.  This is
because,  there is insufficient time to move highly non -relativistic cold
dark matter upto distances greater than the Hubble scale.  Therefore the
entropy per cold dark matter is conserved on supercurvature scales($(k\eta
> 1)$).  So we write:
 \begin{equation}
 \frac{\partial\varepsilon_{b\gamma o}(t)}{\partial t} + 3\varepsilon_{b\gamma o}(t)H(t) =0 
 \end{equation}
The (eq. 63) gives: 
 \begin{equation}
  \varepsilon_{do}\nabla\phi_{o} + \nabla p_{b\gamma} + \frac{\partial\varepsilon_{b\gamma o}}{\partial t}\vec{u_{o}} = 0\end{equation}
 We take the divergence of the above equation  to write:
  \begin{equation}
  6H^{2}\varepsilon_{b\gamma o} = 4\pi G\varepsilon^{2}_{do}
  \end{equation}
  To get the above equation, for near recombination epochs,  we use:
  \begin{equation}
  \nabla^{2}p_{b\gamma}=0\end{equation}
We use the above assumption because for epochs near recombination, the
baryon-radiation plasma starts to decouple fast.  Therefore, only the
gravitational force determines the acceleration of dark matter.  Now for
Adiabatic perturbations, we write:
  \begin{equation}
  \delta S = 0
  \end{equation}
  Also,  we write the following perturbations for other variables:
  \begin{equation}
  \varepsilon_{d} = \varepsilon_{d} = \varepsilon_{do} + \delta\varepsilon_{d}(\vec{x}, t);
  \vec{u} = \vec{u_{o}} + \delta\vec{u}(\vec{x}, t);\phi = \phi_{o} + \delta\phi ;
  p_{b\gamma} = p_{b\gamma o} + \delta p_{b\gamma} = p_{o} + c_{s}^{2}\delta\varepsilon_{b\gamma}
  \end{equation}
Where the variables have their usual meanings (see Sec. ~\ref{sec:2}).  We
use (eq. 71) in (eq. 66) and (eq. 68) to write :
  \begin{equation}
  \frac{\partial\delta\varepsilon_{b\gamma}}{\partial t} = \varepsilon_{b\gamma o}\vec{\nabla}\cdot \delta\vec{u}  + \vec{\nabla}\cdot \lbrace\delta\varepsilon_{b\gamma}\vec{u_{o}}\rbrace = 0
  \end{equation}
 \begin{center} and \end{center}
  \begin{equation}
  \varepsilon_{do}\nabla\delta\phi
  + \delta\varepsilon_{d}\nabla\phi_{o} + \nabla\delta p_{b\gamma o}+\frac{\partial\varepsilon_{b\gamma o}}{\partial t}\delta\vec{u}+ \frac{\partial\delta\varepsilon_{b\gamma}}{\partial t}\vec{u_{o}} = 0
  \end{equation} 
  We use the Langragian co-ordinates and write:
  \begin{equation}
  \lbrace\frac{\partial}{\partial t}\rbrace_{x} =\lbrace\frac{\partial}{\partial t}\rbrace_{q} -\vec{u_{o}}\cdot \nabla_{x}
  \end{equation}
    where \begin{equation}
    \vec{u_{o}} = H(t)q  ; x =aq
    \end{equation}
    We write:\begin{equation}\nabla_{x} = \frac{1}{a}\nabla_{q} ;  \delta = \frac{\delta\varepsilon}{\varepsilon_{o}} 
   \end{equation}
where $\delta$ is the fractional amplitude of  perturbations.  We write
(eq. 75) in the Co-moving coordinates, using (eq. 65) and (eq. 77) to get:
   \begin{equation}
   \dot{\delta_{b\gamma}} + \frac{\vec{\nabla}\cdot \delta\vec{u}}{a} = 0
   \end{equation}
  also we write the (eq. 76) in the co-moving coordinates to get:
  \begin{equation}\frac{\varepsilon_{do}}{a}\nabla\delta\phi + \frac{\varepsilon_{do}}{a}\delta_{d}\nabla\phi_{o} + \frac{c_{s}^{2}}{a}\varepsilon_{b\gamma o}\nabla\delta_{b\gamma}+\dot{\varepsilon_{b\gamma o}}\delta\vec{u}  +\dot{\varepsilon_{b\gamma o}}\delta_{b\gamma}+\varepsilon_{b\gamma o}\dot{\delta_{b\gamma}} -\frac{\vec{u_{o}^{2}}}{a}\varepsilon_{b\gamma o}\vec{\nabla}\cdot \delta_{b\gamma} = 0
   \end{equation}
We write (eq. 69)in co -moving coordinates:
   \begin{equation}
   \dot{\varepsilon_{b\gamma o}} = -\frac{3\varepsilon_{b\gamma o}(t)H(t)}{a} 
   \end{equation}
 from (eq. 71),  we get:
   \begin{equation}
   \varepsilon_{do}\nabla. \delta_{d} = \frac{6H^{2}\varepsilon^{2}_{b\gamma o}\nabla. \delta_{b\gamma}}{8\pi G\varepsilon_{do}}
   \end{equation}
 We take the divergence of (eq. 76), and use (eqn's. 68-77-79) and (eq. 80) to get:
  \begin{equation}
  \frac{12H^{2}\delta_{d}}{a}+ 3aH\dot{\delta_{b\gamma}}+ \lbrace\frac{c_{s}^{2}-u_{o}^{2}}{a}\rbrace\nabla^{2}\delta_{b\gamma} + \nabla. \dot{\delta_{b\gamma}} -\nabla. \delta_{b\gamma}\lbrace\frac{H(2u_{o} + 3)}{a}\rbrace = 0
  \end{equation}
We Neglect the fourth term i. e. $\nabla. \dot{\delta_{b\gamma}}$. This is
because, it is proportional to $\frac{\nabla. \nabla\delta u}{a}$, which is
very small. This is due to very small divergence in the perturbations
to  the ejection velocity of baryon -radiation plasma. With this
assumption, again we argue that the cold dark matter is truly
collisionless. So, in this model, it does not disturb the equilibrium of
the ejecting baryon-radiation plasma. From (eq. 71), we get:
  \begin{equation}8\pi G \varepsilon^{2}_{do}\delta_{d}=6H^{2}\varepsilon^{2}_{b\gamma o}\delta_{b\gamma}
  \end{equation} 
Using the above, we write (eq. 84) as:
  \begin{equation}
   \dot{\delta_{b\gamma}} +\frac{3H^{3}\varepsilon^{2}_{b\gamma}\delta_{b\gamma}}{8\pi Ga^{2}\varepsilon^{2}_{do}}+ \lbrace\frac{c_{s}^{2}- u^{2}_{o}}{3a^{3}H}\rbrace\nabla^{2}\delta_{b\gamma} + \frac{\nabla. \delta_{b\gamma}H^{2}}{3a^{2}}\lbrace2u_{o}+ 3\rbrace = 0  
   \end{equation}
   Therefore, neglecting the term containing $\nabla. \delta_{b\gamma}$, we write:
   \begin{equation}
   \dot{\delta_{b\gamma}} + \frac{3H^{3}\varepsilon^{2}_{b\gamma}\delta_{b\gamma}}{8\pi Ga^{2}\varepsilon^{2}_{do}}+\lbrace\frac{c_{s}^{2} - u^{2}_{o}}{3a^{3}H}\rbrace\nabla^{2}\delta_{b\gamma} = 0
   \end{equation}
 We use: \begin{equation}\delta_{b\gamma}= \int\delta_{b\gamma k}(t)e^{i\vec{k}\cdot \vec{q}}\frac{d^{3}k}{\sqrt{(2\pi)^{3}}}   
 \end{equation}
 Thus,  we write the equation for the evolution  of fractional amplitudes
 of perturbations in the baryon -radiation plasma as:
 \begin{equation}
 \delta_{b\gamma} = \exp\lbrace k^{2}(\frac{c_{s}^{2} - u^{2}_{o}}{3a^{3}H}) -\frac{3H^{3}\varepsilon^{2}_{b\gamma o}}{8\pi Ga^{2}\varepsilon^{2}_{do}}\rbrace t \end{equation} 

For very late  recombination epochs, the radiation starts to decouple
rapidly from matter. Therefore, baryon densities are very low. So we write:
 \begin{equation}
 c^{2}_{s}=\frac{1}{3}
 \end{equation}
 where $c^{2}_{s}$ is the speed of sound. 

The (eq. 89) shows,  that for low baryon densities,   fractional amplitudes
of perturbations in the radiation,  which originate at late recombination
epochs grow at the  fastest rate.  This is because the value of $c_{s}^{2}$
is maximum at late recombination epochs. It is because of the lowest baryon
densities, in the coupled baryon -radiation plasma at these epochs. Also
we can see that the second term in the exponent in (eq. 89)vanishes for
very late recombination epochs, as ratio $\frac{\varepsilon^{2}_{b\gamma
o}}{\varepsilon^{2}_{do}}\rightarrow 0$. We thus write the equation for
evolution of the fractional amplitude of perturbations in the Cold dark
matter, which originate after equality,  when the clustering of dark matter
had already started. 
 \begin{equation}
 \delta_{d} = \frac{3H^{2}}{4\pi G}\frac{\varepsilon^{2}_{b\gamma o}}{\varepsilon^{2}_{do}}\exp \lbrace k^{2}(\frac{c_{s}^{2} - u^{2}_{o}}{3a^{3}H}) -\frac{3H^{3}\varepsilon^{2}_{b\gamma o}}{8\pi Ga^{2}\varepsilon^{2}_{do}}\rbrace t
 \end{equation}
The above equation represents the growth of the fractional amplitudes of perturbations in the cold dark matter,  which originate after equality, for scales $>>H^{-1}$. If baryons contribute  a significant factor of the total matter
density, CDM growth rate will be slowed down between equality and the
recombination epochs \cite{Hu1996}. Also we see that the perturbation
growth rate will slow with scale factor \cite{Caldwell2002}. The
CDM density fluctuations will dominate the density perturbations
of baryon-radiation plasma \cite{Hu1996}. This is because for scales
$>>H^{-1}$, the density perturbations of baryon-radiation plasma are washed
out by the scattering of photons at scales $>\tau_{\gamma}$, which is the
mean free path of photons. The perturbationa in the cold dark matter will
cease to grow when the dark energy starts to dominate \cite{Turner2008}. We
can interpret this from (eq. 91). This is because with the growth of dark
energy, the value of $c_{s}^{2}$ will decrease. This will then lead to
ceasing of growth of CDM perturbations. There is existence of non-linear
structures today. This implies that the growth of fluctuations must have
been driven by non-baryonic dark matter, which was not relativistic at
recombination.  Also,  we see that the perturbations  at supercurvature
scales grow slowly. Recall that these are the modes which enter the
horizon very early, well before the recombination epochs. The slow growth
of such modes is because, it is only at late recombination epochs
that the second term in the exponent in (eq. 91) will vanish due to
$\frac{\varepsilon_{b\gamma o}}{\varepsilon_{do}}\rightarrow 0$. Also, it is
only at late recombination epochs, that the value of $c_{s}^{2}$ reaches
its maximum value of $\frac{1}{3}$, just before decoupling. The amplitude
of the fractional density perturbations in the cold dark matter, in
(eq. 91) will be maximum when the ratio$\frac{\varepsilon_{b\gamma
o}}{\varepsilon_{do}}\rightarrow 1$. This will occur when
 \begin{equation}
  \varepsilon_{do}=\frac{3H^{2}}{2\pi G}
  \end{equation}
  In writing the above equation, we use the result of (eq. 71). So for epochs when the density of baryon-radiation plasma is equal to the density of the cold dark matter, (eq. 91) is:
 \begin{equation}
 \delta_{d} = \frac{\varepsilon_{do}}{2}\exp \lbrace k^{2}(\frac{c_{s}^{2} - u^{2}_{o}}{3a^{3}H}) -\frac{H\varepsilon_{do}}{4a^{2}}\rbrace t
 \end{equation} 

Now we discuss the gravitational instability for scales $<H^{-1}$. This
originates after equality with dark matter dominance in the presence of
strongly coupled baryon - radiation plasma in an expanding universe. We
discuss it in the Newtonian theory. The decoupling of the strongly
coupled baryon - radiation plasma from the non -relativistic cold
dark matter starts after equality. Let us assume that the separation
of this plasma from the cold dark matter gives a relative velocity of
$\vec{u_{o}}$ to the baryon - radiation plasma. At small scales with low
baryon densities, we neglect the contribution of non-diagonal components
in the energy- momentum tensor of dark matter. Therefore, we treat it as
a perfect fluid for many epochs between equality and recombination. We
treat the strongly coupled baryon - radiation plasma as a perfect fluid
for epochs between equality and recombination. This is because of low
baryon -densities at these epochs. Recall that the baryon densities only
starts increasing substantially after recombination. This is when the
primordial nucleosynthesis of hydrogen and helium will start. However, as
an exception in certain regions, the primordial nucleosynthesis may start
at epochs before recombination. We now conclude that there is sufficient
time for the Cold dark matter to flow through distances at scales $<
H^{-1}$. This is because from (eq. 91), we see that the perturbations in the
cold dark matter,  at small scales,  grows at the fastest rate. Therefore
we write the Meshchersky equation as below:
 \begin{equation}
 \varepsilon_{do}\lbrace\frac{\partial\vec{v_{do}}}{\partial t} + (\vec{v_{do}}\cdot \nabla)\vec{v_{do}}\rbrace = -\varepsilon_{do}\nabla\phi_{o}
-\nabla p_{b\gamma}-\frac{\partial\varepsilon_{b\gamma o}}{\partial t}\vec{u_{o}}   
\end{equation}
where \begin{equation}\vec{}u_{o}=\vec{v_{b\gamma o}}- \vec{v_{do}}\end{equation}

At small scales, we assume that in the time that the baryon -radiation
plasma flows out of a given region of space, the inhomogeneity in Cold
dark matter in that time duration is negligible. So we assume that the
baryon-radiation plasma can flow out of a concentrated region of dark
matter without generating the collision terms. This is because the cold
dark matter is highly non-relativistic and collisionless. Also, because
the dark matter has already started to cluster. Therefore we write:
\begin{equation}\vec{v_{b\gamma}}=\vec{v_{b\gamma o}(t)} = H(t)\vec{x}\end{equation}
where $\vec{x}$ is the eulerian cordinate. We write the continuity equation
for flow of the strongly coupled baryon - radiation plasma as :
\begin{equation}
\frac{\partial\varepsilon_{b\gamma}}{\partial t} +\nabla. (\varepsilon_{b\gamma}\vec{v_{b\gamma}})= 0
\end{equation}
Here $\varepsilon_{b\gamma}$ is equal to $\varepsilon_{b\gamma o}(t)$
\begin{equation}
\dot{\varepsilon_{b\gamma o}} = -3H\varepsilon_{b\gamma o}
\end{equation}
We take the divergence of (eq. 94) and neglect the spatial dependence of
$\vec{u_{o}}$.  This is due to the collisionless nature of dark matter. We
also assume that $\varepsilon_{do}$ is constant for small scales. We thus
write the Friedmann equation:
\begin{equation}
\dot{H} + H^{2} = -\frac{4\pi G\varepsilon_{do}}{3}
\end{equation}
For adiabatic perturbations we write from :
   \begin{eqnarray}
\varepsilon_{d}(\vec{x}, t)=\varepsilon_{do} + \delta\varepsilon_{d}(\vec{x}, t);
\vec{v_{d}}(\vec{x}, t)= \vec{v_{do}} + \delta\vec{v_{d}}(\vec{x}, t)= \delta\vec{v_{d}}(\vec{x}, t) ;
\phi(\vec{x}, t)= \phi_{o} +  \delta\phi (\vec{x}, t);\nonumber \\
\vec{u}(\vec{x}, t)= \vec{u_{o}} + \delta\vec{u}(\vec{x}, t);
p_{b\gamma}=p_{b\gamma o}+ c^{2}_{s}\delta\varepsilon_{b\gamma};\delta S=0\end{eqnarray}
where the variables have their usual meanings (see Sec. ~\ref{sec:2}). Then we use the perturbed values of variables in (eq. 94),  to write:
  \begin{eqnarray}
\varepsilon_{do}\delta\dot{\vec{v_{b\gamma}}}-\varepsilon_{do}\dot{\delta\vec{u}}+\delta\varepsilon_{d}\dot{\vec{v_{b\gamma o}}}-\delta\varepsilon_{d}\dot{\vec{u_{o}}} + \varepsilon_{do}(\vec{v_{b\gamma o}}\cdot \nabla)\delta\vec{v_{b\gamma}} -\varepsilon_{do}\vec{u_{o}}\cdot \nabla\delta\vec{v_{b\gamma}} - \varepsilon_{do}\vec{v_{b\gamma o}}\cdot \nabla\delta\vec{u}\nonumber\\
+ \varepsilon_{do}\vec{u_{o}}\cdot \nabla\delta\vec{u}=-\varepsilon_{do}\nabla\delta\phi -\delta\varepsilon_{d}\nabla\phi_{o}-c_{s}^{2}\nabla\varepsilon_{b\gamma}-\dot{\varepsilon_{b\gamma o}}\delta\vec{u} -\delta\dot{\varepsilon_{b\gamma}}\vec{u_{o}}
\end{eqnarray}
   We then use the perturbed values of variables in (eq. 97) to get:
   \begin{equation}
   \delta\dot{\varepsilon_{b\gamma}} +\varepsilon_{b\gamma o}(t)\nabla. \delta\vec{v_{b\gamma}}+
   \vec{v_{b\gamma o}}\nabla. \delta\varepsilon_{b\gamma} =0
   \end{equation}
   We now write (eq. 98) in co-moving coordinates. 
   \begin{equation}
   \dot{\varepsilon_{b\gamma o}} = -\frac{3\varepsilon_{b\gamma o}H}{a}
   \end{equation}
   We also write (eq. 102) in co-moving coordinates to get:
   \begin{equation}
   \dot{\delta\varepsilon_{b\gamma}}=-\frac{\varepsilon_{b\gamma o}}{a}\nabla. \delta\vec{v_{b\gamma}}
   \end{equation}

Using (eq. 95) in (eq. 94) and also using the (eq. 103),  (eq. 104), we write
(eq. 94) in co-moving coordinates. We then take the divergence of the
obtained equation and use the results of (eq. 112) and (eq. 117), to write:
   \begin{eqnarray}\ddot{\delta_{b\gamma}} +H\lbrace 1 + \frac{3}{a} -\frac{u_{o}}{aH}\rbrace\dot{\delta_{b\gamma}} -\lbrace\frac{3\dot{H}}{a}+\frac{3H^{2}\varepsilon_{b\gamma o}}{\varepsilon_{do}}+\frac{9H^{2}\varepsilon_{b\gamma o}}{a\varepsilon_{do}}-\frac{3Hu_{o}}{a^{2}}+4\pi G\varepsilon_{b\gamma o}-\frac{4\pi G\varepsilon_{b\gamma o}}{a}\rbrace\delta_{b\gamma}\nonumber\\ -\frac{3H\varepsilon_{b\gamma o}}{a^{2}\varepsilon_{do}}\lbrace av_{b\gamma o}-1-u_{o}\rbrace\nabla. \delta_{b\gamma}
-\frac{\varepsilon_{b\gamma o}\nabla. \dot{\delta_{b\gamma}}}{a\varepsilon_{do}}+\frac{\varepsilon_{b\gamma o}}{\varepsilon_{do}}\lbrace\frac{c_{s}^{2}-v_{b\gamma o}u_{o}}{a^{2}}\rbrace\nabla^{2}\delta_{b\gamma}=0
\end{eqnarray}
   Also we can write:
\begin{equation}\delta\varepsilon_{b\gamma}=\frac{4c_{s}^{2}\delta\varepsilon_{\gamma}(1-\theta)}{9}\end{equation}
where $\theta=\frac{\delta T}{T}$, represents the anisotropy in the
temperature of radiation. 
The total energy distribution composed of the sum of the dark matter
and the baryon-radiation plasma is a function of time in an expanding
universe. If this remains smooth inspite of the inhomogeneities at scales
smaller than $H^{-1}$, we can write:
\begin{equation}
\varepsilon_{b\gamma}a^{4}(t) + \varepsilon_{d}a^{3}(t)= E_{total}(t)\end{equation}
or for any time $'t'$, we can write:
\begin{equation}-a\delta\varepsilon_{b\gamma}= \delta\varepsilon_{d} \end{equation} 
At epochs near recombination, when the baryon-radiation plasma is decoupled
from the cold dark matter,  we assume that their bulk velocity does
not appreciably change in these epochs.  Therefore, it remains constant
for epochs near recombination. The ejection velocity of baryon-radiation
plasma is affected only due to the gravitational force of the dark matter
potential. There is no force due to the pressure of the baryon-radiation
plasma. This can be either due to low pressure of the plasma, due to
higher baryonic densities, or due to some new force of strange energy(dark
energy), having negative pressure. The baryon densities can never be very
high before recombination. This is because it would hinder future hydrogen
nucleosynthesis after recombination. So, it would be rightly concluded that
a new form of dark energy having negative pressure starts to dominate
at these epochs. So, we write:
 \begin{equation}
 \vec{v_{b\gamma o}}=\text{constant}\Longrightarrow\dot{\vec{v_{b\gamma}}}=0 \end{equation}
 \begin{equation}\dot{\vec{v_{do}}}=\nabla\phi_{o}\end{equation}
The above equation comes from the same argument, which we give in
(eq. 36). The actual velocity of the baryon-radiation plasma stops to
increase, when it is almost about to decouple from dark matter. At these
epochs, the acceleration of dark matter will reverse its sign.  Using the
above equation and (eq. 95), we write:
 \begin{equation} \dot{\vec{u_{o}}}=-\nabla\phi_{o}\end{equation}
  and in co-moving coordinates,  \begin{equation}\dot{\vec{u_{o}}}=-\frac{\nabla\phi_{o}}{a}\end{equation}
  \begin{equation}\nabla. \delta\dot{u} =-\nabla\delta\phi\end{equation} 
  and in co-moving coordinates, 
  \begin{equation}\nabla. \delta\dot{u} =-\frac{\nabla\delta\phi}{a}\end{equation}
 or \begin{equation}\nabla. \delta\dot{u}=-\frac{\nabla^{2}\delta\phi}{a} =-\frac{4\pi G\delta\varepsilon_{d}}{a}\end{equation}
Using (eq. 103), we can write the above equation as:
\begin{equation}
\nabla. \delta\dot{u}=4\pi G\varepsilon_{b\gamma o}\delta_{b\gamma}\end{equation}
 Also we assume, 
 \begin{equation}\nabla. \dot{\delta u}=0\end{equation}

For scales much smaller than the Jeans length$\lambda < \tau_{\gamma}<
\lambda_{J}$ i. e.  scales much smaller than the curvature scale i. e for
which $H^{-1}$ is dominant, we neglect the terms $\propto H$ and write
Eq. $105$ as:
 \begin{equation}
 \ddot{y_{k}}-\frac{u_{o}}{a}\dot{y_{k}}+\lbrace 4\pi G\varepsilon_{b\gamma o}(1-\frac{1}{a})-k^{2}\varepsilon_{b\gamma o}(\frac{c_{s}^{2}-v_{b\gamma o}u_{o}}{a^{2}})-\frac{\dot{u_{o}}}{a}\rbrace y_{k}=0
 \end{equation}
 We use (eq. 39)and ((eq. 40) in writing the above equation. If we now choose: 
 \begin{eqnarray}
 2\lambda = -\frac{u_{o}}{a}\nonumber \\
 \lbrace 4\pi G\varepsilon_{b\gamma o}(1-\frac{1}{a})-k^{2}\varepsilon_{b\gamma o}(\frac{c_{s}^{2}-v_{b\gamma o}u_{o}}{a^{2}})-\frac{\dot{u_{o}}}{a}\rbrace =\omega^{2}
  \end{eqnarray}
 Then the auxillary equation is:
 \begin{equation}
 D^{2}+2\lambda D + \omega^{2}=0
 \end{equation}
 with
 \begin{equation}
 D=-\lambda \pm \sqrt{\lambda^{2}-\omega^{2}}
 \end{equation}
 The only solution which is of physical significance is when $\lambda <
 \omega$, i. e.  when the roots of the auxillary equation are imaginary, i. e. :
 \begin{eqnarray}
 D=-\lambda + i\alpha \nonumber \\ 
 \alpha^{2}=\omega^{2}-\lambda^{2}\nonumber \\
 y_{k}=A\sqrt{\lbrace 1 + (\frac{\lambda}{\alpha})^{2}\rbrace}e^{-\lambda t}\cos\lbrace\alpha -\arctan\frac{\lambda}{\alpha}\rbrace
 \end{eqnarray} 
 
 Thus we write:
 \begin{equation}
\delta\varepsilon_{\gamma}\lbrace 1 - \Theta\rbrace =A\sqrt{\lbrace 1 + (\frac{\lambda}{\alpha})^{2}\rbrace}e^{-\lambda t}\cos\lbrace\alpha -\arctan\frac{\lambda}{\alpha}\rbrace
 \end{equation}
for $\lambda < \tau_{\gamma}< \lambda_{J}$,  we can write for free -streaming photons:
\begin{equation}
 \frac{1}{p}\frac{dp}{dt} = -H - \frac{\partial\phi}{\partial t} - \frac{p^{i}}{a}\frac{\partial\psi}{\partial x^{i}}\end{equation}
 where $\psi$ corresponds to the Newtonian potential in Perturbed FRW
 universe. Here we assume,   \begin{equation}\phi = \psi\end{equation}
 This is because the potential is dominated by the cold dark matter
 between epochs of equality and recombination.   We neglect the spatial
 inhomogeneities in the Newtonian potential at very small scales
 and  terms $\propto H$.  Using the same argument that we give for
 (eq. 56. ), we get:
 \begin{equation}
 \delta\varepsilon_{\gamma}= \frac{1}{3}\delta p = -\frac{e^{-\phi_{do}}\delta\phi_{d}}{3}
 \end{equation}
 Using (eq. 123)and (eq. 126), we write:
  \begin{equation}
  \frac{-e^{-\phi_{do}}\delta\phi_{d}\lbrace 1 -\Theta\rbrace}{3} =A\sqrt{\lbrace 1 + (\frac{\lambda}{\alpha})^{2}\rbrace}e^{-\lambda t}\cos\lbrace\alpha -\arctan\frac{\lambda}{\alpha}\rbrace 
  \end{equation}
  \begin{center}or
  \end{center}
  \begin{equation}
  \lbrace\Theta -1\rbrace = \frac{3A e^{\phi_{do}}}{\delta\phi_{d}}\sqrt{\lbrace 1 + (\frac{\lambda}{\alpha})^{2}\rbrace}e^{-\lambda t}\cos\lbrace\alpha -\arctan\frac{\lambda}{\alpha}\rbrace
  \end{equation}

This equation is of same form as the one we get for Einstein's static
universe in Jeans theory. In the static universe,  we arrived at this
equation by an assumption of (eq. 32). This we reasoned,   may be partly
possible where baryon densities are higher due to early primordial
synthesis of Helium before recombination. But the primary reason would
be that dark energy, with negative pressure starts to dominate from
these epochs. The similarity in form of the two equations,  shows that
the epochs, for which the  CMB temperature anisotropy derived from the
solutions in the static Einstein universe and those from the expanding
Newtonian match,  will be the epochs when the dark energy starts to
dominate. We see that the amplitude in (eq. 123)is variable, but at epochs
when the temperature anisotropy of radiation is dominant, which is the
late recombination epochs, the term $\lambda$ can be negleted. So we write
(eq. 123) as:
  \begin{equation}
  \lbrace\Theta -1\rbrace =\frac{3A e^{\phi_{do}}}{\delta\phi_{d}}\cos\alpha
  \end{equation}
  Using the result of (eq. 61), we write:
  \begin{equation}
  \theta\approx\frac{\delta\phi_{d}}{3}
  \end{equation}

The equation for fractional amplitudes  of perturbations in the cold
dark matter that originate at scales smaller than the mean free path of
the photons is:
  \begin{equation}
  \delta_{d} = -\frac{4Ac_{s}^{2}a}{9\varepsilon_{do}}\sqrt{\lbrace 1 + (\frac{\lambda}{\alpha})^{2}\rbrace}e^{-\lambda t}\cos\lbrace\alpha -\arctan\frac{\lambda}{\alpha}\rbrace
  \end{equation}
  and for late recombination epochs, neglecting $\lambda$,  we can write:
  \begin{equation}
  \delta_{d} = -\frac{4Ac_{s}^{2}a}{9\varepsilon_{do}}\cos\alpha 
    \end{equation}
 where \begin{equation} \alpha =\sqrt{\lbrace 4\pi G\varepsilon_{b\gamma o}(1-\frac{1}{a})-k^{2}\varepsilon_{b\gamma o}(\frac{c_{s}^{2}-v_{b\gamma o}u_{o}}{a^{2}})-\frac{\dot{u_{o}}}{a}\rbrace -(\frac{-u_{o}}{2a})^{2}}
  \end{equation}

The above equation shows that the fractional amplitudes of perturbations
in cold dark matter at small scales after equality keep growing with
scale factor, but with a decreasing frequency of oscillations. 

\section{Conclusion}
\label{conclusion}
 
We conclude that this model, wherein we assume that the baryon -radiation
plasma has an ejection velocity,  flowing out of regions of highly
non- relativistic cold dark matter,  explains the effect of cold dark
matter perturbations on CMB temperature anisotropy fluctuations fairly
well. The results of (eq. 62) and (eq. 130) represent the Sachs-Wolfe
effect. It also correctly predicts the growth of fractional amplitudes
of perturbations in cold dark matter, which originate between equality
and recombination epochs.  The (eq. 59) and (eq. 127) describe the affect
of the perturbations in the cold dark matter on the CMB temperature
anisotropy fluctuations. They also highlight that the dominant component
of CMB temperature anistropy fluctuations  is at red-shifts of very late
recombination epochs $z\approx 1100$. This is because the perturabtions
in the cold dark matter are minimal at these epochs. The unperturbed
potential $\phi_{do}$ remains frozen at epochs between equality and
recombination. Therefore  the fluctuations in the CMB  temperature
anisotropy between epochs of equality and recombination are determined by
the perturbations in the potential of cold dark matter potential only. The
assumptions of (eq. 32) and (eq. 111), assume the growth of a strange form
of energy(dark energy), with negative pressure, which cancels the force
due to the pressure of baryon-radiation plasma. 

These assumptions also
represent the fact that the rate of increase of ejection velocity of
the Baryon -radiation plasma is only due to the gravitational field
of the Cold dark matter.  It can be inferred that the force due to
its pressure can vanish only to some extent due to low pressure in
high baryonic density regions. This is due to early primordial helium
nucleosynthesis in such regions. Since  the baryonic densities can never
be very high before recombination. This is because it will hinder the
hydrogen nucleosynthesis which starts after recombination. Therefore, it
has to be assumed that the primary reason for the vanishing of pressure
of baryon-radiation plasma, at near recombination epochs, is that the
dark energy term with negative pressure starts to dominate at these
epochs. The epochs when the solutions of the (eq. 59) and (eq. 127)
match will give the correct time scales when the Dark energy starts
to dominate. This work can be taken up in a following paper. The use of
(eq. 27) for strongly coupled baryon -radiation plasma indicates its
equilibrium at all epochs between equality and recombination,  even in
the presence of dark matter. This assumtion can only be made if the dark
matter is truly collisionless. The correct explanation of pridominant
contribution to CMB temperature anisotropy from very late recombination
epochs, and the Sachs-Wolfe effect, based on the assumptions of (eq. 27), 
emphasizes the truly collisionless nature of cold dark matter.

Thus it
can be concluded that the use of this model for describing the dynamics
of Cold dark matter,  in presence of strongly coupled Baryon -radiation
plasma for epochs between equality and recombination is sufficiently
useful. More accurate results can be obtained by studying the dynamics
of baryon -radiation plasma using energy momentum tensor of an imperfect
fluid in such a scenario. This will be attempted in a future work. 

\begin{acknowledgements} 
We wish to thank Md. Mahfoozul Haque for valuable advice in formatting of
the manuscript in Latex. Also, we thank Zaheer and Divyendu for valuable
discussions and reading the manuscript. 
\end{acknowledgements}

   \end{document}